\documentclass[twocolumn,showpacs,preprintnumbers,amsmath,amssymb]{revtex4}

\usepackage{amsmath,amssymb}
\usepackage{graphicx}
\usepackage{dcolumn}
\usepackage{bm}

\usepackage{graphicx,color}

\def\Beq{\begin{equation}}
\def\Eeq{\end{equation}}

\def\Bea{\begin{eqnarray}}
\def\Eea{\end{eqnarray}}

\def\Beaa{\begin{eqnarray*}}
\def\Eeaa{\end{eqnarray*}}

\def\BD{\begin{description}}
\def\ED{\end{description}}

\def\BC{\begin{center}}
\def\EC{\end{center}}

\def\Bcenter{\begin{center}}
\def\Ecenter{\end{center}}

\def\del{\partial}
\def\<{\langle}
\def\>{\rangle}
\def\({\left(}
\def\){\right)}

\def\d{\delta}
\def\D{\Delta}

\def\t{\tau}
\def\s{\sigma}

\def\^{\wedge}

\input{epsf.tex}

\begin{document}

\title{Asymptotic principal values and 
regularization methods for   correlation functions with reflective boundary conditions}

\author{Masafumi Seriu}
 \email{mseriu@edu00.f-edu.u-fukui.ac.jp}
 \affiliation{%
 Department of Physics, Graduate School of Engineering, 
University of Fukui, 
Fukui 910-8507, Japan\\}%


\begin{abstract}

We introduce a concept of asymptotic principal values which enables us 
to handle rigorously singular integrals of higher-order poles encountered in the computation of  various quantities based on correlation functions of a vacuum.

Several theorems on asymptotic 
principal values are proved and they are expected to become  bases for investigating and developing 
some class of regularization methods for singular integrals.

We make use of these theorems for analyzing  mutual relations between some  regularization methods, including 
a  method naturally derived from asymptotic principal values.  
It turns out that the concept of asymptotic principal values and the theorems for them are quite useful in this type of analysis, 
providing a suitable language to describe  what is discarded and what is retained in each regularization method. 
\end{abstract}

\maketitle

\section{\label{sec:Introduction}Introduction}

Physics of quantum vacuum fluctuations is one of the 
intriguing research topics expected to  be developed 
through the  interplay between theories, experiments and practical applications. 

Investigations of quantum vacuum fluctuations even stimulate the border area 
between physics and mathematics. 
As a typical example of this sort, we often encounter singular integrals in computing several quantities based on 
correlation functions of a vacuum in question. 
The occurrence of singularity or divergence is often  a signal of 
surpassing the border of validity  of a model by too much extrapolation. Furthermore, it could originate from 
deeper physical processes for  which   satisfactory consistent mathematics is still unavailable. 
How to handle singular integrals  can be  then a challenging topic, requiring  both mathematical analysis and physical considerations.

Faced with singular integrals, we need to resort to some regularization method to get a finite result.
The aim of this paper is to give an organized mathematical basis underlying some typical regularization methods and 
make clear their mutual relations. We introduce below a concept of {\it asymptotic principal values} which can be a
key tool to analyze some class of regularization methods. We then prove several theorems on 
the asymptotic principal values useful for studying regularization procedures. 

There are still various uncertainties to clear up in regularization methods,
 reflecting our lack of mathematical basis for  handling infinities. 
 In this situation, we cannot expect any universal regularization method, but we need to 
 customize the method by try and error depending on the problem in question. 
It is far from  the aim of this paper to judge which method is better than the others. 
It just tries to present a concrete mathematical basis for further  considerations  and developments of  better regularization methods.

The organization of the paper is as follows. In Sec. \ref{sec:Example}, we present one 
simple example where singular integrals of a higher-order pole emerge. The origin of singularities in this example is physically 
clear and we can get some idea on how these integrals should be regularized. 
In Sec. \ref{sec:Theorems}, we introduce {\it asymptotic principal values} which describe precisely 
how singular integrals behave,  which 
 can be useful for  investigating and developing several regularization methods. 
We then prove several useful theorems on the asymptotic principal values. 
In Sec. \ref{sec:Regularization}, we analyze some typical regularization methods 
by means of theorems prepared in the previous section. 
Sec. \ref{sec:SummaryDiscussions} is devoted for a summary and several discussions.

\section{\label{sec:Example}Typical example of singular integrals}

Let us consider as an example the measurement of  the electromagnetic vacuum fluctuations in a half-space bounded by 
a perfectly reflecting infinite mirror. 
Recently the switching effect~\cite{Switching} and the smearing effect 
due to the quantum spread of a probe particle~\cite{Smearing} have been analyzed 
by studying the measurement process of the Brownian particle released in this environment.

We thus take a  model introduced in Ref.\cite{YuFord} and reanalyzed in Ref.'s \cite{Switching} and \cite{Smearing}: 
 Suppose that a flat, infinitely spreading mirror of perfect reflectivity is 
 placed on the $xy$-plane  ($z=0$). Then let us investigate   the quantum vacuum fluctuations of the electromagnetic field 
 inside  the half-space $z>0$ 
 by  releasing a classical charged probe-particle with mass $m$ and charge $e$ in the environment.  
We can estimate  the quantum  fluctuations of the vacuum through 
the velocity dispersions of the probe-particle released in the environment.

  When the velocity of the probe-particle is much smaller than the light velocity $c$, 
  the  motion for the particle is described by
\Beq
 m  \frac{d\vec{v}}{dt}= e \vec{E}(\vec{x},t)\ \ , 
 \label{eq:eqofmotion}
\Eeq
where  $\vec{E} (\vec{x},t)$ is the electric field. 

Within the time-period when  the particle does not move so much, 
Eq. (\ref{eq:eqofmotion}) along with the initial condition $\vec{v}(0)=\vec{v}_0$ is solved approximately as
\Beq
   \vec{v}(t) \simeq  \vec{v}_0 + \frac{e}{m}\  {\int_0}^t \vec{E}(\vec{x},t') dt'\ \ . 
 \label{eq:velocity}
\Eeq

For simplicity, let us consider only the ``sudden-switching" case; 
the measurement is switched on  abruptly, stably continued  for $\t$ [sec] before  switched off abruptly. 
It is  mathematically described by a step-like switching function without any switching tails. 
The velocity dispersions of the particle, $\< {\D v_i}^2 \>$ ($i=x, y, z$), are then given by 
\Beq
\< {\D v_i}^2 (\vec{x}, \t)\> = 
\frac{e^2}{m^2} \int_0^\t dt' \int_0^\t dt'' 
\< E_i (\vec{x} , t') E_i (\vec{x} , t'')  \>_R \ \ , 
\label{eq:v2sudden}
\Eeq
by noting that 
$\< E_i (\vec{x} , t)  \>_R=0$: Here $\< E_i (\vec{x} , t') E_i (\vec{x} , t'')  \>_R$ ($i=x, y, z$) are 
 the renormalized two-point correlation functions of the electric field (the suffix ``R" is for ``renormalized").
 Now $\< E_i (\vec{x} , t') E_i (\vec{x} , t'')  \>_R$ ($i=x, y, z$) are 
computed~\cite{BroMac} as 
\Bea
 \< E_z (\vec{x} , t') E_z (\vec{x} , t'')  \>_R 
&=& \frac{1}{\pi^2} \frac{1}{ (T^2 - (2z)^2)^2} 
\label{eq:EzEz} \\
 \< E_x (\vec{x} , t') E_x (\vec{x} , t'')  \>_R 
&=& \< E_y (\vec{x} , t') E_y (\vec{x} , t'')  \>_R \nonumber  \\
&& = - \frac{1}{\pi^2} \frac{T^2 + 4z^2}{ (T^2 - (2z)^2)^3} \ \ , 
\label{eq:ExEy}
\Eea
where   $T:= t'-t''$. 
(We set $c=\hbar=1$ hereafter throughout the paper.) 

It is obvious that the integral in Eq.(\ref{eq:v2sudden}) is regular when $\t < 2z$, but 
 singular when $\t > 2z$, reflecting the singularity 
at $T=2z$ inherent in the correlation functions $\< E_i (\vec{x} , t') E_i (\vec{x} , t'')  \>_R$ 
given in Eqs.(\ref{eq:EzEz}) and (\ref{eq:ExEy}). 

For the present purpose, it suffices to show only the result of $\< {\D v_z}^2 \>$ for $\t>2z$~\cite{Switching}, 
\Bea
&& \< {\D v_z}^2 \> \\
&&\ \  = 
  \frac{e^2}{32 \pi^2m^2}  
  \left\{
 \frac{\t}{z^3}\ln\left( \frac{\t+2z}{\t-2z} \right)^2 + \frac{8(1-\frac{2z}{\t})}{z^2 \rho} + O(\rho) 
 \right\}
 \nonumber \\
 &&\ \  \sim \frac{e^2}{4 \pi^2 m^2 z^2} \left(1+ \frac{1}{\rho}\right)  \ \  ({\rm for}\ \t \gg 2z)\ \ , 
\label{eq:Dv2M}
\Eea
where $\rho$ $(>0)$ is a dimension-free asymptotic parameter for handling 
the singular integral properly (see the next section for details). 
Accordingly  the above expression 
should be understood as an asymptotic expression as $\rho \sim 0$. 

This result is derived by a formula for an {\it asymptotic principal value}, 
the rigorous definition of which shall be given in the next section; 
\Bea
&& \wp_{(\rho)} \int_0^1 dx\  \frac{1-x}{(x^2 - \s^2)^2} \nonumber \\
&& \qquad \qquad  = \frac{1}{8\s^3}\ln\left( \frac{1+\s}{1-\s} \right)^2 +  
 \frac{1- \s}{ 2\s^2 \rho} + O(\rho) \ \ ,  
\label{eq:useful}
\Eea
for $0< \s <1$.  One can derive Eq.(\ref{eq:useful}) with the help of {\sf Theorem 1}; 
a more direct  derivation is also found in {\it Appendix C}  of  Ref.\cite{Switching}. 

Leaving the rigorous treatment of singular integrals for the next section, 
we here focus on the physical reason why the singularity of correlation functions occurs at $|T|=2z$.  
Due to the mirror-reflections of 
signals with the light velocity, the values of the electric field at 
the two world-points $(t', x, y, z)$ and $(t'' , x, y, z)$ 
are expected to be strongly correlated when $|t' -t''|=2z$. 
When the measuring time $\t$ is short enough (shorter than the travel-time of the signal $2z$), then it always follows $|t' -t''|< 2z$, so that 
 these correlations are not captured by the probe. When the measuring time is long enough ($\t > 2z$), however, 
 these strong correlations accumulate in  the velocity fluctuations of the particle at $z$.  
Therefore it is expected that the resulting  singular term of the form $A/\rho$ ($A>0$) 
contains information on the reflecting boundary. 

On the other hand,  typical  regularization procedures\cite{DaviesDavies} correspond to  discarding  
such a  singular term (e.g. the $1/\rho$ term in Eq.(\ref{eq:Dv2M})) in effect. 
It should be clarified when this type of regularization is valid and when not. 
We shall discuss on this point in more detail in Sec.{\ref{sec:Regularization}}. 

It turns out that the model given here is too simplified and should be modified taking into account the 
switching effect\cite{Switching} and the smearing effect due to the quantum spread of the 
probe-particle\cite{Smearing}. However, 
it suffices for the present purpose of giving some example of singular integrals.

\section{\label{sec:Theorems} Basic formulas for handling  singular integrals}

In view of the example in the previous section, it is clear that we sometimes need to 
estimate a singular integral whose integrand possesses a higher-order pole.
In order to investigate various regularization methods later,  we first need some concrete quantity 
corresponding  to  a singular integral 
for which all the information is retained and nothing is discarded. 
Then, the following asymptotic 
definition of a singular integral may be relevant. 
\vskip .2cm
\underline{\it Definition 1} 
\vskip .2cm

Let $f(x)$ be an arbitrary real function defined around an interval $[a,b]$, differentiable at $x=c$ ($a<c<b$) sufficiently many times. 
For a positive integer $n$,  then, let us introduce an {\it asymptotic principal value} of order $\rho$ defined by
\Bea
\wp_{(c, \rho)} (f,n)&:=& \wp_{(\rho)} \int_a^b \frac{f(x)}{(x-c)^n}\ dx \nonumber \\
               &:=& \left\{\int_a^{c-\rho} + \int_{c+\rho}^b \right\} \frac{f(x)}{(x-c)^n}\ dx\ \ ,
\label{eq:PV}
\Eea
where $\rho$ is a sufficiently small positive parameter. $\Diamond$ 

The asymptotic principal value is a generalization of 
the standard Cauchy principal value, corresponding to 
$\lim_{\rho \downarrow 0}{\wp}_{(c, \rho)} (f,1)$, 
 in two ways. First,  
the order of singularity $n$ can be greater than 1. Second, 
only the asymptotic behavior as $\rho \sim 0$ is concerned and 
the convergence for the limit $\rho \rightarrow 0$ is not necessarily required.  
In other words, we focus on how the integral  behaves near $\rho \sim 0$ rather than 
the $\rho \rightarrow 0$ limit itself. In this sense, 
all the information is retained and no infinities are discarded in defining the asymptotic principal value.

Let us now   introduce another asymptotic quantity: 
\vskip .2cm
\underline{\it Definition 2} 
\vskip .2cm
With the same premises as in {\it Definition 1}, we define 
\Beaa
\wp_{(\rho)} \left[ \frac{f(x)}{(x-c)^n} \right]_a^b 
&:=& \left[ \frac{f(x)}{(x-c)^n} \right]_a^{c-\rho}+\left[ \frac{f(x)}{(x-c)^n} \right]_{c+\rho}^b \  .\ \   \Diamond
\Eeaa

It is easily shown that 
\Bea
\wp_{(\rho)} \left[ \frac{f(x)}{(x-c)^n} \right]_a^b 
&=&  \left[ \frac{f(x)}{(x-c)^n} \right]_a^b - \frac{1}{\rho^n} \{ f \}_{(n)}^{(c,\rho)} 
\label{eq:SBracket2}
\Eea
with 
\Bea
\{ f \}_{(n)}^{(c,\rho)}&:=& f(c+\rho) -(-)^n f(c-\rho) \ \ .
\label{eq:f-factor}
\Eea

By a Taylor-expansion in $\rho$, it is obvious 
that $\{ f \}_{(n)}^{(c,\rho)}$ is $O(\rho)$ for an even $n$ (when $f'(c)\neq 0$), so that 
the term $\frac{1}{\rho^n} \{ f \}_{(n)}^{(c,\rho)}$ in  Eq.(\ref{eq:SBracket2}) is $O(1/\rho^{n-1})$ and 
singular  in the $\rho \rightarrow 0$ limit. 
Similarly, $\{ f \}_{(n)}^{(c,\rho)}$ is  $O(1)$ for an odd $n$ (when $f(c)\neq 0$), so that 
$\frac{1}{\rho^n} \{ f \}_{(n)}^{(c,\rho)}$ is  $O(1/\rho^{n})$ and singular as  $\rho \rightarrow 0$. 

The following definition is just for making formulas below concise. 
\vskip .2cm
\underline{\it Definition 3} 
\vskip .2cm

\Bea
\zeta_{(c,\rho)}(f,n)&:=& \frac{1}{n}\ 
\wp_{(\rho)} \left[ \frac{f(x)}{(x-c)^n} \right]_a^b \  . \ \ \Diamond
\label{eq:zeta}
\Eea

With these preparations, let us start with the following lemma:
\vskip .5cm
\fbox{{\sf Lemma 1} }
\vskip .5cm
For any function $f(x)$ differentiable at $x=c$ and for any integer $n$ ($\geq 2$), it follows that 
\Bea
&& {\wp}_{(c, \rho)} (f,n)= \frac{1}{n-1} \wp_{(c, \rho)}(f',n-1) - \zeta_{(c,\rho)}(f,n-1)\ \ . \nonumber \\
&& 
\label{eq:formula1}
\Eea
\vskip .5cm
\underline{{\sl Proof}} : 
\vskip .5cm
Noting that 
\[
\frac{1}{(x-c)^n}=-\frac{1}{n-1} \left( \frac{1}{(x-c)^{n-1}} \right)'\ \ ,
\]
we have 
\Bea
&& {\wp}_{(c, \rho)}(f,n) := -\frac{1}{n-1} \wp_{(\rho)} \int_a^b \left( \frac{1}{(x-c)^{n-1}} \right)' f(x)\ dx \nonumber \\
                       &&=  -\frac{1}{n-1} \left\{
                       \wp_{(\rho)} \left[  \frac{f(x)}{(x-c)^{n-1}} \right]_a^b \right. \nonumber \\
                       && \qquad \qquad \qquad \qquad \qquad \qquad \left.    -\wp_{(\rho)} \int_a^b \frac{f'(x)}{(x-c)^{n-1}}\ dx
                           \right\} , \nonumber \\
&& 
\label{eq:RigorousPartialInt}
\Eea
where the partial-integral has been performed to get the last line. Then the equality follows.  $\Box$

\vskip .5cm
We now prove a formula which relates a multi-pole integral with a simple-pole integral: 
\vskip .5cm
\fbox{{\sf Theorem 1} }
\vskip .5cm
For any function $f(x)$ differentiable sufficiently many times at $x=c$ and for any integer $n$ ($\geq 2$), it follows that 
\Bea
&& {\wp}_{(c, \rho)} (f,n)    \nonumber \\
&=& \frac{1}{(n-1)!}\, \wp_{(c, \rho)}(f^{(n-1)},1) \nonumber \\
&& \qquad  - \sum_{k=1}^{n-1}\frac{(n-k)!}{(n-1)!}\, \zeta_{(c,\rho)}(f^{(k-1)},n-k)\ \ .
\label{eq:formula2}
\Eea
\vskip .5cm
\underline{{\sl Proof}} : 
\vskip .5cm
($1^\circ$) 
For  $n=2$, 
the claimed equality reduces to  
\[
\wp_{(c, \rho)} (f,2)= \wp_{(c, \rho)}(f',1) - \zeta_{(c,\rho)}(f,1)\ \ , 
\]
which clearly holds due to {\sf Lemma 1}. 

($2^\circ$) 
Let us assume that the equality holds for a function $F(x)$ and for  $n=m$ ($m \geq 2$), i.e., 
\Bea
&& \wp_{(c, \rho)} (F,m)  \nonumber \\
&=& \frac{1}{(m-1)!}\, \wp_{(c, \rho)}(F^{(m-1)},1) \nonumber   \\
&& \ \ - \sum_{k=1}^{m-1}\frac{(m-k)!}{(m-1)!}\, \zeta_{(c,\rho)}(F^{(k-1)},m-k)\ \ .
\label{eq:n=m}
\Eea
Now applying {\sf Lemma 1} for $n=m+1$, we have 
\Beaa
&& \wp_{(c, \rho)} (f,m+1) = \frac{1}{m}\, \wp_{(c, \rho)} (f', m) - \zeta_{(c,\rho)}(f , m) \\
  &=& \frac{1}{m}\, \left\{ 
  \frac{1}{(m-1)!}\, \wp_{(c, \rho)}(f^{(m)},1)  \right. \\
  && \left. - \sum_{k=1}^{m-1}\frac{(m-k)!}{(m-1)!}\, \zeta_{(c,\rho)}(f^{(k)},m-k)
    \right\}
    - \zeta_{(c,\rho)}(f , m)\ \ , 
\Eeaa
where the assumed equation (\ref{eq:n=m}) for  $F(x)=f'(x)$  has been used to get the second equality.
Rearranging the summation, the last equality reduces to 
\Beaa
&& \wp_{(c, \rho)} (f,m+1)  \nonumber \\
&=& \frac{1}{m!}\, \wp_{(c, \rho)}(f^{(m)},1) \nonumber   \\
&& \ \ - \sum_{k=1}^{m}\frac{((m+1)-k)!}{m!}\, \zeta_{(c,\rho)}(f^{(k-1)},(m+1)-k)\ \ .
\Eeaa
Thus the claimed equation (\ref{eq:formula2}) holds for $n=m+1$.

($3^\circ$) By the mathematical induction, Eq.(\ref{eq:formula2}) holds for any integer $n$ ($n \geq 2$). $\Box$

\vskip .3cm
Based on {\sf Theorem 1}, it is natural  to introduce a quantity $\tilde{\wp}_{(c, \rho)} (f,n)$, which is 
a simple-pole part plus a regular part of 
$\wp_{(c, \rho)} (f,n)$, putting aside  singular contributions  from higher-order poles:   
\vskip .2cm
\underline{\it Definition 4} 
\vskip .2cm
We  define 
\Bea
&&\tilde{\wp}_{(c, \rho)} (f,n):=
 \frac{1}{(n-1)!}\, \wp_{(c,\rho)} (f^{(n-1)},1) \nonumber \\
&& \qquad  - \sum_{k=1}^{n-1}\frac{(n-k-1)!}{(n-1)!}
 \left[\frac{f^{(k-1)}(x)}{(x-c)^{n-k}} \right]_a^b\ \ \ .\ \  \Diamond \nonumber 
\Eea

The ``mild part" $\tilde{\wp}_{(c, \rho)} (f,n)$ of $\wp_{(c, \rho)} (f,n)$   
shall be important in the discussion of  regularization methods in Sec.{\ref{sec:Regularization}} below. 

We now have a formula which enables us to separate singular contributions from a multi-pole integral:
\vskip .5cm
\fbox{{\sf Theorem 2} }
\vskip .5cm
For any function $f(x)$ differentiable sufficiently many times at $x=c$ and for any integer $n$ ($\geq 2$), it follows that 
\Bea
&& \wp_{(c, \rho)} (f,n)  = \tilde{\wp}_{(c, \rho)} (f,n) \nonumber \\
&&\qquad   +  \sum_{k=1}^{n-1}\frac{(n-k-1)!}{(n-1)!}\, \frac{1}{\rho^{n-k}} \{ f^{(k-1)} \}_{(n-k)}^{(c,\rho)} \ \ .
\label{eq:formula3}
\Eea
\vskip .5cm
\underline{{\sl Proof}} : 
\vskip .5cm
It is straightforward to show this formula  due to {\sf  Theorem 1} along with Eqs.(\ref{eq:zeta}) and (\ref{eq:SBracket2}). $\Box$

\vskip .5cm
\fbox{{\sf Lemma 2} }
\vskip .5cm
For any function $f(x)$ differentiable  at $x=c$ and for a positive integer $n$, it follows that 
\Bea
\wp_{(c, \rho)} (f,n+1) = 
 \frac{1}{n}\, {\del_c}\, \wp_{(c,\rho)} (f,n) + \frac{1}{n \rho^n}\left\{ f \right\}_n^{(c,\rho)} \ \ .
\label{eq:del-formula0}
\Eea
\vskip .5cm
\underline{{\sl Proof}} : 
\vskip .5cm
We compute ${\del_c}\, \wp_{(c,\rho)} (f,n)$ directly as  
\Beaa
&& {\del_c}\, \wp_{(c,\rho)} (f,n)=
\del_c \left(\left\{\int_a^{c-\rho} + \int_{c+\rho}^b \right\} \frac{f(x)}{(x-c)^n}\ dx \right) \\
&& = n \wp_{(c,\rho)} (f,n+1) -\frac{1}{\rho^n} \{ f \}_n^{(c,\rho)}\ \ ,
\Eeaa
where the second term in the last line comes from the $c$-derivative applied to 
the upper- and the lower-limit of 
the integral region. Thus the claimed equation follows. $\Box$

\vskip .5cm
\fbox{{\sf Theorem 3} }
\vskip .5cm
For any function $f(x)$ differentiable sufficiently many times at $x=c$ and for an integer $n$ ($\geq 2$), it follows that 
\Bea
\tilde{\wp}_{(c, \rho)} (f,n) = 
 \frac{1}{(n-1)!}\, {\del_c}^{n-1}\, \wp_{(c,\rho)} (f,1) \ \ .
\label{eq:del-formula}
\Eea
\vskip .5cm
\underline{{\sl Proof}} : 
\vskip .5cm
Due to {\sf Theorem 2}, the claimed equation (\ref{eq:del-formula}) is equivalent to 
\Bea
&& \wp_{(c, \rho)}(f,n) = 
 \frac{1}{(n-1)!}\, {\del_c}^{n-1}\, \wp_{(c,\rho)} (f,1) \nonumber \\
&& \qquad  +  \sum_{k=1}^{n-1}\frac{(n-k-1)!}{(n-1)!}\, \frac{1}{\rho^{n-k}} \{ f^{(k-1)} \}_{(n-k)}^{(c,\rho)} \ \ . 
\label{eq:del-formula2}
\Eea
Thus it suffices to  show Eq.(\ref{eq:del-formula2}). 

($1^\circ$) 
Let us consider the case $n=2$, where 
the R.H.S. (right-hand side) of Eq.(\ref{eq:del-formula2}) becomes 
\Beaa
&& \frac{\del}{\del c} \left(\left\{\int_a^{c-\rho} + \int_{c+\rho}^b \right\} \frac{f(x)}{x-c}\ dx \right) \\
&& \qquad \qquad \qquad                    + \frac{1}{\rho}\left\{f(c-\rho) + f(c+\rho) \right\} \ \ . \\
\Eeaa
In this expression, the 
$c$-derivative applied to the upper- and the lower-limit of 
the integral region 
 yields a term which exactly cancels the second term. As a result, the above expression reduces to 
$\wp_{(c, \rho)} (f,2)$, i.e., the L.H.S. (left-hand side) of Eq.(\ref{eq:del-formula2}).     
Thus  Eq.(\ref{eq:del-formula2}) holds for $n=2$.

($2^\circ$) 
Let us now assume that Eq.(\ref{eq:del-formula2}) holds for  $n=m$ ($m \geq 2$), i.e., 
\Bea
&& \wp_{(c, \rho)}(f,m) = 
 \frac{1}{(m-1)!}\, {\del_c}^{m-1}\, \wp_{(c,\rho)} (f,1) \nonumber \\
&& \qquad  +  \sum_{k=1}^{m-1}\frac{(m-k-1)!}{(m-1)!}\, \frac{1}{\rho^{m-k}} \{ f^{(k-1)} \}_{(m-k)}^{(c,\rho)} \ \ . 
\label{eq:n=m2}
\Eea
Due to {\sl Lemma 2}, then, it becomes
\Beaa
&& \wp_{(c, \rho)} (f,m+1) = \frac{1}{m} \del_c \wp_{(c,\rho)} (f,m) 
+ \frac{1}{m\rho^m} \{ f \}_{m}^{(c,\rho)}  \\
&&=\frac{1}{m} \del_c \left\{ 
 \frac{1}{(m-1)!}\, {\del_c}^{m-1}\, \wp_{(c,\rho)} (f,1) \right.  \\
&& \qquad \left.  +  \sum_{k=1}^{m-1}\frac{(m-k-1)!}{(m-1)!}\, \frac{1}{\rho^{m-k}} \{ f^{(k-1)} \}_{(m-k)}^{(c,\rho)} 
\right\} \\
&& \qquad \qquad + \frac{1}{m\rho^m} \{ f \}_{m}^{(c,\rho)}  \ \ ,
\Eeaa
where Eq.(\ref{eq:n=m2}) has been used to get the last line. 
Noting that the relation 
\[
\del_c \{ f \}_{m}^{(c,\rho)}=\{ f' \}_{m}^{(c,\rho)}\ \ , 
\]
which obviously holds from Eq.(\ref{eq:f-factor}), it reduces to 
\Beaa
&& \wp_{(c, \rho)} (f,m+1)
 = \frac{1}{m!}\, {\del_c}^m\, \wp_{(c,\rho)} (f,1)  \\
&& \qquad  +  \sum_{k=1}^{m}\frac{(m-k)!}{m!}\, \frac{1}{\rho^{m-k+1}} \{ f^{(k-1)} \}_{(m-k+1)}^{(c,\rho)} \ \ . 
\Eeaa
Thus Eq.(\ref{eq:del-formula2}) holds for $n=m+1$.

($3^\circ$) By the mathematical induction, Eq.(\ref{eq:del-formula2}) holds for any integer $n$ ($n \geq 2$). 
Thus the claimed formula Eq.(\ref{eq:del-formula}) has been shown. 
 $\Box$

\section{\label{sec:Regularization} Typical regularization methods and their mutual relations}

Based on the results in the previous section, let us now come back to 
the problem of regularization methods for singular integrals. 

Let us consider a typical singular integral 
\Beq
I=\int_a^b \frac{f(x)}{(x-c)^n}\ dx\ \ 
\label{eq:Iexample}
\Eeq
for any function $f(x)$ differentiable sufficiently many times at $x=c$ ($a < c < b$) 
and for a positive integer $n$.

\subsection{\label{sec:PartialIntegralMethod} Regularization method with  partial integrals}

The first method of regularization we consider is a  {\it method with partial integrals} which is sometimes 
made used of. 
We insert an identity 
\[
\frac{1}{(x-c)^n}=-\frac{1}{n-1} \left( \frac{1}{(x-c)^{n-1}} \right)'\ \ ,
\]
into Eq.(\ref{eq:Iexample}) and {\it formally} perform a partial integral:
\Bea
&& I= -\frac{1}{n-1} \int_a^b \left( \frac{1}{(x-c)^{n-1}} \right)' f(x)\ dx \nonumber \\
                       &&=  \frac{1}{n-1} \left\{
                       \int_a^b \frac{f'(x)}{(x-c)^{n-1}}\ dx
                       - \left[  \frac{f(x)}{(x-c)^{n-1}} \right]_a^b
                            \right\} . \nonumber \\
&& 
\label{eq:NaivePartialInt}
\Eea
In this way, the order of singularity is reduced by one. Repeating the 
similar procedure, the integral $I$ is reduced to the $n=1$ case for which 
the prescription of the Cauchy principal value  may be applied.

Due to {\sf Theorem 2}, however, it is obvious that 
singular terms should exist and should have been  discarded by hand in the above procedure. 
Indeed, compared with Eq.(\ref{eq:NaivePartialInt}) with the  rigorous 
expression Eq.(\ref{eq:RigorousPartialInt}), it is obvious 
that the singularities which should reside in the second term on the R.H.S. of 
Eq.(\ref{eq:NaivePartialInt}) are simply discarded by hand. 
Thus, in view of {\sf Theorem 2},  the above method is equivalent to the simple replacement of  $I$ as
\Bea
I  \mapsto \tilde{\wp}_{(c,\rho)} (f,n)\ \ .
\label{eq:NaivePartialIntReplace}
\Eea

There is still  room, however, to regard the method of partial integrals as 
a shorthand prescription of what we here call the  {\it method of 
infinitesimal imaginary part}, which is much more of theoretical grounds~\cite{DaviesDavies}.
We shall consider this method in the next subsection.

\subsection{\label{sec:ImaginaryMethod} Regularization method with  infinitesimal imaginary part}

The method of infinitesimal imaginary part is based on  well-known Dirac's formula~\cite{Dirac} for  an  integral kernel, 
\Bea
\frac{1}{(x-c) \pm i \rho}= \wp_{(c,\rho)} \frac{1}{x-c} \mp i \pi \d(x-c) \ \ , 
\label{eq:Dirac-formula}
\Eea
which is most easily shown by estimating  an integral 
$\int_a^b \frac{f(x)}{(x-c) \pm i \rho}$ by means of an appropriate contour-integral 
for a suitable function $f(x)$. 

By differentiating  the both-sides of Eq.(\ref{eq:Dirac-formula}) $n-1$ times 
with respect to $c$, and by applying 
{\sf Theorem 3}, 
we get 
\Bea
\frac{1}{\left((x-c) \pm i \rho \right)^n}= \tilde{\wp}_{(\rho,c)} (\ \cdot \ , n) \mp i \frac{\pi}{(n-1)!} \d^{(n-1)}(x-c) \ \ .
\label{eq:Dirac-formula2}
\Eea
in the sense of  an integral kernel. 
Recalling {\it Definition 4}, however, we see that the R.H.S. is reduced to 
the $n=1$ case, for which the prescription of the Cauchy principal value may be applied. 

It is notable that just the introduction of some infinitesimal imaginary part results
 in a tamable quantity 
such as $\tilde{\wp}_{(\rho,c)} (\ \cdot \ , n)$ at the cost of the imaginary contribution of the second term on the 
R.H.S. of Eq.(\ref{eq:Dirac-formula2}). Thus along with some causality arguments~\cite{DaviesDavies}, it is often argued  that 
the singular integral $I$ should be interpreted as the real part of 
$\int \frac{f(x)}{\left((x-c) \pm i \rho \right)^n} dx$, i.e., 
\Bea
I \mapsto  \Re \int \frac{f(x)}{\left((x-c) \pm i \rho \right)^n} dx\ \ . 
\label{eq:InterpretReal}
\Eea

As far as one is evaluating  {\it real} quantities, one may further argue that the second term on the R.H.S. of Eq.(\ref{eq:Dirac-formula2}) shall not 
contribute. If so, the procedure is in effect equivalent to the replacement Eq.(\ref{eq:NaivePartialIntReplace}). 
In this sense, the method of partial integrals discussed in the previous subsection may be justified provided that  it is 
regarded as a shorthand prescription of the method of infinitesimal imaginary part.

Another way of looking at this method is to pay attention to the L.H.S. (rather than the R.H.S.) of 
Eq.(\ref{eq:Dirac-formula2}). 
As far as computations of real quantities are concerned, then,  
this method is equivalent to the replacement 
$\frac{1}{(x-c)^n}$ with 
$\left(\frac{x-c}{(x-c)^2+\rho^2} \right)^n$ along with  taking the limit $\rho \rightarrow 0$ after evaluating the integral:
\Bea
I \mapsto  \lim_{\rho \rightarrow 0} \int_a^b f(x)  \left(\frac{x-c}{(x-c)^2+\rho^2} \right)^n dx\ \ . 
\label{eq:InterpretReal2}
\Eea

There is some subtle points in this method. 
One of them is to  discard the imaginary part of the R.H.S. of Eq.(\ref{eq:Dirac-formula2}) 
on the grounds that one is evaluating  {\it real} quantities. Considering that 
the regularization has been achieved at the cost of introducing  the imaginary part though tiny,  
the imaginary part should carry important information and some concern naturally arises whether one can discard it so freely. 

Indeed, a simple example can be presented for which this kind of procedure fails. 
Let us consider  an integral $I_1= \int_{-1}^1 dx$ which is purposefully regarded as 
\[
I_1=\int_{-1}^1 x \cdot \frac{1}{x} dx\ \ .
\]
It is obvious that $I_1=2$.  The analysis by the asymptotic principal value (see the next subsection) 
also results in $I_1^{(\rho)} \rightarrow 2$ in the limit $\rho \rightarrow 0$. 
This is because all the  information is retained  in the prescription of the asymptotic principal value.

On the other hand, the above mentioned scheme makes a replacement  
\[
\frac{1}{x} \mapsto \frac{x}{x^2+\rho^2}=\frac{1}{2}\left(\ln (x^2+\rho^2) \right)'\ \ , 
\]
so that 
\Bea
I_1 \rightarrow && I_1^{(\rho)}:=\int_{-1}^1 x \cdot \frac{1}{2}\left(\ln (x^2+\rho^2) \right)' dx  \nonumber \\
              && = \ln (1+\rho^2) -\int_0^1 \ln (x^2 + \rho^2) dx\ \ ,
\label{eq:CounterEx}
\Eea
where a partial integral has been performed to get the last line. 
However, it is clear that $I_1^{(\rho)}\rightarrow \infty $ as $\rho \rightarrow 0$, contradicting 
with the obvious result $I_1=2$.

Quite interestingly, no contradiction occurs for 
$I_m=\int_{-1}^1 x^m \cdot \frac{1}{x} dx$ with $m\geq 2$ since the second term 
in Eq.(\ref{eq:CounterEx}) 
becomes $-m \int_0^1 x^{m-1} \ln (x^2 + \rho^2) dx$ so that no singularity occurs around $x\sim 0$ for $m\geq 2$. 
More generally,  the integral of the form $\int_{-1}^1 \frac{f(x)}{x} dx $, if treated by the above prescription,  gives rise to 
the dominant contribution $-f'(0) \int_0^1 \ln (x^2 + \rho^2) dx $ which diverges as $\rho \rightarrow 0$. 

With these caveats in mind, let us now 
move to a new regularization method based on  the asymptotic principal values.

\subsection{\label{sec:Rho-Method} Regularization method with  asymptotic principal values}

Let us finally introduce a new 
regularization method based on the  asymptotic principal values. 

For  the simple example in  Sec.\ref{sec:Example}, there has been  a definite 
physical interpretation of 
the singularity in the correlation function. Furthermore, 
the system considered there has been a combination of quantum objects 
with a macroscopic mirror. Therefore it might  be  also probable that the deepest cause of the singularity 
resides in  the validity issue  of the  model originating from too much  extrapolation from the quantum side to the macroscopic 
situation. 
Indeed there is an investigation showing that 
the quantum fluctuations of the mirror boundary   drastically decrease  the singular behavior 
near the mirror~\cite{FordSvaiter}. Therefore it is reasonable to take the origin of the singularity more 
realistically (rather than just mathematical phenomenon), expecting   that some physical processes suppress 
the order of singularity. 

Going back to  the example of the integral $I$ in Eq.(\ref{eq:Iexample}), then, it is possible to 
interpret  $I$ in the sense of an asymptotic principal value, 
\Beq
I \mapsto \wp_{(c, \rho)} (f,n)\ \ ,
\label{eq:AsymptoticMethod}
\Eeq
with 
the dimension-free parameter $\rho$ being  provided  by the ratio 
of  some natural cut-off scale with the system-size in question. (For instance, 
the ratio of the plasma wave-length of the mirror with $2z$ for the example 
in Sec.\ref{sec:Example}).  
The advantage of this regularization scheme  is that one can explicitly analyze the 
$\rho$-dependence of the 
integral. For instance, one may study the influence of the quantum fluctuations of the mirror by 
treating $\rho$ as a fluctuation parameter. 
The result for $\< {\D v_z}^2 \>$ given in Eq.(\ref{eq:Dv2M}) along with Eq.(\ref{eq:useful}) 
is an example of the computation by the 
method of asymptotic principal values.

We see that {\sf Theorem 2} is the basis for understanding   the relation between the regularization methods 
discussed so far.  The difference between the method of asymptotic principal 
value ($\wp_{(c, \rho)} (f,n)$) and the method of infinitesimal imaginary part 
($\tilde{\wp}_{(c, \rho)} (f,n)$) is given by the second term on the R.H.S. of 
Eq.(\ref{eq:formula3}), which is of $O(1/\rho^{n-1})$.

\section{\label{sec:SummaryDiscussions} Summary and discussions}

In this paper, we have focussed on  singular integrals with a higher-order pole 
which frequently emerge in computing quantities based on 
two-point correlation functions of a vacuum. 

To deal with this type of singular integrals, we have introduced the concept of  
{\it asymptotic principal values}.
The asymptotic principal value of order $\rho$,  which is a generalization of the Cauchy principal 
value, is defined by 
 introducing a cut-off parameter $\rho$, focussing  solely on 
the  asymptotic behavior of the integral as $\rho \sim 0$. In this sense, it is a rigorous 
object retaining all the information on the singular integral. 

We have then proved several theorems on asymptotic principal values  which are expected to  serve as bases for studying regularization methods for singular integrals. 

To see how asymptotic principal values can be made use of, 
 we have selected  three typical regularization methods and have 
analyzed  their mutual relations with the help of theorems we have prepared.  It has turned out that the concept of asymptotic principal values 
and related theorems are quite useful in this kind of analysis. Indeed, in terms of 
asymptotic principal values, it has been possible to  describe 
without ambiguity  what is discarded and what is retained in each regularization method. 

No universal  regularization method is available so far and we need to carefully select or invent a 
suitable method depending on the problem in question. 
For instance, we recall the example in Sec.\ref{sec:Example} where velocity dispersion of the probe, 
$\< {\D v_z}^2 \>$, is sensitive to the  regularization method. 
In particular, the result expected by the method of infinitesimal imaginary part 
(Sec.\ref{sec:ImaginaryMethod}) (and the method  of partial integrals 
(Sec.\ref{sec:PartialIntegralMethod})) is 
\[
\< {\D v_z}^2 \> \sim \frac{e^2}{4 \pi^2 m^2 z^2}\ \ ({\rm for} \ \t \gg 2z) .
\]
On the other hand, 
the result expected by the method of asymptotic principal values (Sec.\ref{sec:Rho-Method}) is 
\[
\< {\D v_z}^2 \> \sim \frac{e^2}{4 \pi^2 m^2 z^2} \left(1+ \frac{1}{\rho}\right)\ \ ({\rm for} \ \t \gg 2z)\ \ ,
\]
by choosing $\rho$ in the order of  the ratio of plasma wave-length and the typical size $2z$. 
Strictly speaking, the model in Sec.\ref{sec:Example} is a too simplified one and should be modified taking into account the 
quantum spread of the probe-particle itself. Then the behavior of $\< {\D v_z}^2 \>$ at late time
is corrected to a more reasonable one $\< {\D v_z}^2 \> \sim 1/\t^2$ rather than $\sim 1/z^2$
~\cite{Smearing}. 

In any case it is significant  to compare the results derived by different regularization methods 
in more detail for approaching to   a more 
satisfactory mathematical theory of  regularization procedures.

\section*{Acknowledgement}
The author would like to thank C. H. Wu for various  discussions.

\vskip .5cm


\begin{thebibliography}{99}
\bibitem{Switching}
M. Seriu and C. H. Wu, Phys. Rev. {\bf A77}, 022107 (2008).
\bibitem{Smearing}
M. Seriu and C. H. Wu, 
Phys. Rev. {\bf A80}, 052101 (2009).
\bibitem{YuFord}
H. Yu and L. H. Ford, Phys. Rev. {\bf D70}, 065009 (2004).
\bibitem{BroMac}
L. S. Brown and G. J. Maclay, Phys. Rev. {\bf 184}, 1272 (1969).
\bibitem{DaviesDavies}
e.g., K. T. R. Davies and R. W. Davies, Can. J. Phys. {\bf 67}, 759 (1989).
\bibitem{Dirac}
P. A. M. Dirac, {\sl The Principles of Quantum Mechanics} (4th Edition) (Oxford University Press 1958) \S50  Eq.(35).
\bibitem{FordSvaiter}
L. H. Ford and N. F. Svaiter, Phys. Rev. {\bf D58}, 065007 (1998). 
\end{thebibliography}
\end{document}